\documentclass[aps,tightlines]{revtex4}

\usepackage{amsfonts}
\usepackage{amsmath}
\usepackage{amssymb}
\usepackage{graphicx}

\begin{document}

\title[Entanglement generation...]{Entanglement generation and transfer
between remote atomic qubits interacting with squeezed field}
\author{Paulo Jos\'{e} dos Reis}
\email{paulojreis@uel.br}
\affiliation{Departamento de F\'{\i}sica, Universidade Estadual de Londrina, Londrina
86051-990, PR Brazil }
\author{S. Shelly Sharma}
\email{shelly@uel.br}
\affiliation{Departamento de F\'{\i}sica, Universidade Estadual de Londrina, Londrina
86051-990, PR Brazil }
\author{N. K. Sharma}
\email{nsharma@uel.br}
\affiliation{Departamento de Matem\'{a}tica, Universidade Estadual de Londrina, Londrina
86051-990 PR, Brazil }
\thanks{}

\begin{abstract}
A pair of two level atoms $A_{1}A_{2}$, prepared either in a separable state
or in an entangled state, interacts with a single mode of two mode squeezed
cavity field while a third atomic qubit $B$ interacts with the second mode
of the squeezed field in a remote cavity. We analyze, numerically, the
generation, sudden death and revival of three qubit entanglement as a
function of initial entanglement of qubits $A_{1}A_{2}$ and degree of
squeezing of electromagnetic field. Global negativity of partially
transposed state operator is used to quantify the entanglement of three atom
state. It is found that the initial entanglement of two mode field as well
as that of the pair $A_{1}A_{2}$ , both, contribute to three atom
entanglement.  A maximally entangled single excitation Bell pair in first
cavity and two mode field with squeeze parameter $s=0.64$ are the initial
conditions that optimize the peak value of three qubit mixed state
entanglement. A smaller value of $s=0.4$ under similar conditions is found
to generate a three qubit mixed state with comparable entanglement dynamics
free from entanglement sudden death.
\end{abstract}

\maketitle


\section{Introduction}

Quantum entanglement is an essential physical resource in communication
protocols \cite{eker91} and information processing \cite{bouw00}. Remote
quantum systems may become entangled through interaction with a third
quantum system as in cavity QED experiments \cite{raim01}, where
entanglement of photon state in two cavities results from interaction with
an atomic qubit. On the other hand, atomic qubits having shared quantized
motion may be entangled through interaction with electromagnetic field as in
ion traps \cite{turc98}. A parametric down converter is known to generate
two mode electromagnetic field in a squeezed state with the entanglement of
two modes determined by squeeze parameter. Direct observation of 10 dB
squeezing of quantum noise of light, has been reported recently \cite{vahl08}%
. As such, two mode squezed states are a potential entanglement resource. It
has been shown that the field state entanglement can be transferred to a
pair of remote atoms \cite{son02,pate04} or three remote atoms \cite{reis09}
distributed in two isolated cavities. Paternostro et al., \cite{pate09} have
investigated the connection between entanglement-transfer to a pair of
non-interacting two-level systems and statistical properties of entangled
two-mode continuous variable resource. State engineering through bilinear
interactions between two remote qubits and two-mode Gaussian light fields
has, also, been reported \cite{ades10}. An interesting study of entanglement
transfer from a three-mode quantized field to a system of three spatially
separated qubits, each one made of a two-level atom resonantly coupled to a
cavity mode has been presented in ref. \cite{casa09}. Entanglement transfer
from freely propagating quantized light to an atomic system has been
achieved experimentally \cite{juls04,eisa05,chan05}. In this article,
distributed tripartite entanglement generation through entanglement transfer
from two mode squeezed field to three atomic qubits in two cavities, is
proposed. Two level atoms, $A_{1}$ and $A_{2}$, prepared either in a
separable state or in an entangled state interact with a single mode of two
mode squeezed field in a cavity held by Alice, while a third atom $B$
located in a remote cavity interacts with the second mode of the squeezed
field. We analyze, analytically and numerically, the entanglement dynamics
of atomic qubits after tracing over the field degrees of freedom. This is a
natural way of obtaining a distributed channel for quantum communication,
when the entangled resource is the continuous variable (CV) state of a
photonic system.

Global negativity \cite{zycz98,eise99,vida02} and $K-$way negativities \cite%
{shar07} are used to qualify and quantify the free entanglement of three
qubit mixed state. In our earlier article \cite{reis09}, the three atom
system in separable initial state was found to develop $W-$like entanglement
exhibiting entanglement sudden death (ESD) and entanglement sudden revival
(ESR). Entanglement sudden death, reported by Yu and Eberly \cite{yu04,yu06}
for the first time, refers to disappearance of entanglement in finite time.
ESD observed experimentally for entangled photon pairs \cite{alme07}, and
atomic ensembles \cite{laur07} is a hindrance to using the system for
implementing useful protocols. It is known that entanglement can be
distilled from a three qubit pure or mixed state having free entanglement.
With this in mind, we look for initial conditions on three atom state and
squeezed field state so as to reduce the time interval between ESD and ESR
or make the phenomenon disappear altogether. The three qubit entanglement
generation, sudden death and revival dynamics depends strongly on degree of
squeezing of two mode vacuum state and initial entanglement of pair $%
A_{1}A_{2}$. It is found that the initial state quantum correlations of
qubits $A_{1}A_{2}$ translate into an increase in remote qubit entanglement
and a remarkable change in the rate at which the entanglement decays.

The paper is organized as follows. A brief description of the model and
procedure to obtain analytical expressions for three atom state at current
time from different initial states of atoms and field are given in Section
II. Section III focuses on a comparative analysis of remote qubit
entanglement dynamics for different initial states. A summary of results is
presented in section IV.

\section{The Model}

We consider the entanglement transfer process from a two-mode squeezed
vacuum field to a system of three localized and spatially distributed
qubits. An entangled two-mode CV state is generated with an off-line
process. Two atomic qubits, $A_{1}$ and $A_{2}$, localized inside a single
mode cavity $c_{1}$ interact resonantly with one mode of the field. A third
two level atom $B$, located in cavity $c_{2}$, interacts with the second
field mode. We assume that each mode of the squeezed field is first injected
into a cavity and then interacts resonantly with atomic qubits. The scheme
used is analogous to that of our earlier work in which all three atomic
qubits are prepared, initially, in their ground states. Here we generalize
the model to investigate the effect of initial two qubit entanglement on
tripartite entanglement generation. A search for squeeze parameter value and
initial two qubit entangled state, suitable for generating three qubit
correlations that overcome entanglement sudden death, is carried out. The
action of two-mode squeezing operator%
\begin{equation}
\hat{S}(s)=\exp (-s\widehat{a}\widehat{b}+s\widehat{a}^{\dagger }\widehat{b}%
^{\dagger }),
\end{equation}%
on two-mode vacuum state $\left\vert 0,0\right\rangle $ generates two-mode
squeezed vacuum state 
\begin{equation}
\left\vert \Psi _{F}\right\rangle =\frac{1}{\cosh s}\overset{\infty }{%
\underset{n=0}{\sum }}(\tanh s)^{n}\left\vert n,n\right\rangle ,
\end{equation}%
where $\widehat{a}^{\dagger }$, $\widehat{a}$ and $\widehat{b}^{\dagger }$, $%
\widehat{b}$ are the bosonic creation and annihilation operators for modes
one and two, respectively. The two mode squeezed state is an entangled state
having bipartite entanglement determined by value of squeeze parameter $s$.
The variances of quadrature operators for $\left\vert \Psi _{F}\right\rangle 
$ are below the vacuum limit. Simple linear coupling is used to inject one
of the field modes from a two mode field source into cavity $c_{1}$ in
vacuum state, while the second field mode is directed to the remote cavity c$%
_{2}$ in vacuum state. Neglecting cavity mode dissipation, the resonant
cavity-CV field mode coupling is described by beam splitter operator 
\begin{equation}
\hat{B}_{i}(\theta )=\exp \left[ -\frac{\theta }{2}(\hat{f}_{i}^{\dagger }%
\hat{c}_{i}-\hat{f}_{i}\hat{c}_{i}^{\dagger })\right] ,\quad i=1,2,
\label{1}
\end{equation}%
where $\hat{c}_{i}$ $\left( \hat{f}_{i}\right) $and $\hat{c}_{i}^{\dagger
}\left( \hat{f}_{i}^{\dagger }\right) $are creation and annihilation
operators for photons inside the i$^{th}$cavity (external field),
respectively. The coupling between the cavity field and the external field
is determined by the cavity mirror transmittance coefficient $T(\theta
)=\cos ^{2}\left( \frac{\theta }{2}\right) $. After injecting the two-mode
non-classical field into independent cavities $c_{1}$ and $c_{2}$, the
cavity field at $t=0$ is found to be in a mixed state 
\begin{eqnarray}
\widehat{\rho }_{F}(0) &=&\left( \frac{1}{\cosh s}\right) ^{2}\overset{%
\infty }{\underset{n,m=0}{\sum }}\underset{k,l=0}{\overset{\min \left[ n,m%
\right] }{\sum }}(\tanh s)^{n+m}G_{kl}^{nm}(\theta )  \label{fieldzero} \\
&&\times \left( \left\vert n-k\right\rangle _{c_{1}}\left\langle
m-k\right\vert \right) \left( \left\vert n-l\right\rangle
_{c_{2}}\left\langle m-l\right\vert \right) ,
\end{eqnarray}%
where

\begin{equation}
G_{kl}^{nm}(\theta )=C_{k}^{n}\left( \theta \right) C_{k}^{m}\left( \theta
\right) C_{l}^{n}\left( \theta \right) C_{l}^{m}\left( \theta \right) ,
\end{equation}%
and

\begin{equation}
C_{k}^{n}\left( \theta \right) =\sqrt{\frac{n!}{k!(n-k)!}}\cos ^{k}\frac{%
\theta }{2}\sin ^{n-k}\frac{\theta }{2}.  \label{cnk}
\end{equation}%
The degree of entanglement of $\widehat{\rho }_{F}(0)$ is determined by the
transmittance coefficient $T(\theta )$ and is maximal at $T(\theta )=1$. The
beam splitter has a disentangling effect. The composite field $\widehat{\rho 
}_{F}(0)$ is in a mixed state for $\cos \theta <1$, while when $\cos \theta
=\sin \theta $ it is in a separable state.

\subsection{Atom Field Interaction}

Consider $N$ identical two level atoms interacting via dipole coupling with
a single-mode quantized radiation field in a resonator. The ground and
excited states for the atom $i$ ($i=1\ $to $N$) are, respectively, denoted
by $\left\vert g\right\rangle _{i}$ and $\left\vert e\right\rangle _{i}$.
Spin operators for $i^{th}$ atomic qubit are defined as $\widehat{\sigma }%
_{z}^{i}=\left\vert e\right\rangle _{i}\left\langle e\right\vert -\left\vert
g\right\rangle _{i}\left\langle g\right\vert $, $\widehat{\sigma }%
_{-}^{i}=\left\vert g\right\rangle _{i}\left\langle e\right\vert $ and $%
\widehat{\sigma }_{+}^{i}=\left\vert e\right\rangle _{i}\left\langle
g\right\vert $. Defining collective spin variables of $N$ two-level atoms as 
$\widehat{\sigma }_{k}=\sum_{i=1,N}\widehat{\sigma }_{k}^{i}$ where $%
k=(z,+,-)$, we may construct the eigenbasis of operators $\widehat{\sigma }%
^{2}$ and $\widehat{\sigma }_{z}$ to represent $N$ atom internal states. A
typical basis vector in coupled basis is written as $\left\vert \sigma
,m_{\sigma }\right\rangle $, with eigen values of $\widehat{\sigma }^{2}$
and $\widehat{\sigma }_{z}$ given by $\sigma \left( \sigma +2\right) $ and $%
m_{\sigma }$, respectively.

In the absence of atom-field coupling, the free Hamiltonian given by $%
\widehat{H}_{at}=\frac{\hbar \omega _{a}}{2}\widehat{\sigma }_{z}$ for
atomic qubits and $\widehat{H}_{cav}=\hbar \omega _{0}\left( \widehat{a}%
^{\dagger }\widehat{a}\right) $ for cavity field, determines the system
dynamics. Here $\hbar \omega _{a}$ is the level splitting of the two-level
atoms, $\omega _{0}$ is a frequency of the electromagnetic field and $%
\widehat{a}^{\dagger }$($\widehat{a}$) is photon creation (annihilation)
operator. The atom-field interaction Hamiltonian given by Tavis Cummings
model (TCM) \cite{tavi78} in interaction picture and rotating wave
approximation, has the form%
\begin{equation}
\widehat{H}_{int}=\hbar g_{c}\underset{i=1,N}{\sum }\left( \widehat{a}%
\widehat{\sigma }_{+}^{i}+\widehat{a}^{\dagger }\widehat{\sigma }%
_{-}^{i}\right) ,  \label{hint}
\end{equation}%
for resonant ($\omega _{0}=\omega _{a}$) interaction of dipoles with cavity
field. Here $g_{c}$ is the atom-field coupling strength assumed to be the
same for all atoms. Since the Hamiltonian commutes with $\left( \widehat{%
\sigma }\right) ^{2}$, the unitary evolution operator $\widehat{U}_{1}(\tau
)=\exp \left[ \frac{-i}{\hbar }\widehat{H}t\right] $ conserves the quantum
number $\sigma $.

For two atoms the coupled basis vectors are the set of symmetric states $%
\left\vert 2,-2\right\rangle $, $\left\vert 2,0\right\rangle $, $\left\vert
2,2\right\rangle $ and antisymmetric state $\left\vert 0,0\right\rangle $.
The number state of cavity field is represented by $\left\vert
n\right\rangle $. For two atoms in cavity $c_{1}$, interacting resonantly $%
\left( \omega _{0}-\omega _{a}=\delta =0\right) $ with $n$ photons,
interaction hamiltonaian is represented by matrix%
\begin{equation*}
H_{int}=\left[ 
\begin{array}{ccc}
\hbar \omega _{0}n & \hbar g\sqrt{2n} & 0 \\ 
\hbar g\sqrt{2n} & \hbar \omega _{0}n & \hbar g\sqrt{2\left( n-1\right) } \\ 
0 & \hbar g\sqrt{2\left( n-1\right) } & \hbar \omega _{0}n%
\end{array}%
\right] ,
\end{equation*}%
in the basis $\left\vert 2,-2,n\right\rangle $, $\left\vert
2,0,n-1\right\rangle $, and $\left\vert 2,2,n-2\right\rangle $. Using
eigenvalues and eigenbasis of $H_{int}$, the unitary operator that governs
the evolution of two atoms in cavity c$_{1}$ is found to be

\begin{equation}
\text{$U_{1}^{n}(\tau )=\exp ^{-i\omega _{0}nt}\left( 
\begin{array}{ccc}
\frac{B_{n}^{2}\cos (f_{n}\tau )+A_{n}^{2}}{A_{n}^{2}+B_{n}^{2}} & \frac{%
-iB_{n}\sin (f_{n}\tau )}{\sqrt{\left( A_{n}^{2}+B_{n}^{2}\right) }} & \frac{%
A_{n}B_{n}\left[ \cos (f_{n}\tau )-1\right] }{A_{n}^{2}+B_{n}^{2}} \\ 
\frac{-iB_{n}\sin (f_{n}\tau )}{\sqrt{\left( A_{n}^{2}+B_{n}^{2}\right) }} & 
\cos (f_{n}\tau ) & \frac{-iA_{n}\sin (f_{n}\tau )}{\sqrt{\left(
A_{n}^{2}+B_{n}^{2}\right) }} \\ 
\frac{A_{n}B_{n}\left[ \cos (f_{n}\tau )-1\right] }{A_{n}^{2}+B_{n}^{2}} & 
\frac{-iA_{n}\sin (f_{n}\tau )}{\sqrt{\left( A_{n}^{2}+B_{n}^{2}\right) }} & 
\frac{\left[ A_{n}^{2}\cos (f_{n}\tau )+B_{n}^{2}\right] }{%
A_{n}^{2}+B_{n}^{2}}%
\end{array}%
\right) $},  \label{u_two}
\end{equation}%
where interaction parameter $\tau =gt$, $f_{n}=$ $\sqrt{2\left( 2n-1\right) }
$, $A_{n}=\sqrt{2\left( n-1\right) }$, and $B_{n}=\sqrt{2n}$. For a single
atom the basis states $\left\vert 1,-1\right\rangle $ $\left( \left\vert
1,1\right\rangle \right) $ stands for the ground (exited) state of the atom.
The unitary matrix that determines the state evolution due to interaction of
a single atom with field in cavity $c_{2}$, reads as

\begin{equation}
\text{$U_{2}^{m}(\tau )$}=\left( 
\begin{array}{cc}
\cos \left( \sqrt{m}\tau \right) & -i\sin \left( \sqrt{m}\tau \right) \\ 
-i\sin \left( \sqrt{m}\tau \right) & \cos \left( \sqrt{m}\tau \right)%
\end{array}%
\right) ,  \label{u_one}
\end{equation}%
in the basis $\left\vert 1,-1,m\right\rangle $, $\left\vert
1,1,m-1\right\rangle $. The evolution operator for the two cavity composite
system is obtained by taking the tensor product that is

\begin{equation}
U_{12}^{nm}(\tau )=\text{$U_{1}^{n}(\tau )\otimes U_{2}^{m}(\tau ).$}
\label{eq11}
\end{equation}

For a given atomic initial state%
\begin{equation}
\widehat{\rho }_{A_{1}A_{2}B}(0)=\left\vert \Phi \right\rangle \left\langle
\Phi \right\vert =\left( \left\vert \Phi _{A_{1}A_{2}}(0)\right\rangle
\left\langle \Phi _{A_{1}A_{2}}(0)\right\vert \right) _{c_{1}}\left(
\left\vert \Phi _{B}(0)\right\rangle \left\langle \Phi _{B}(0)\right\vert
\right) _{c_{2}}  \label{atomzero}
\end{equation}%
and field state $\widehat{\rho }_{_{F}}(0)$ (Eq. (\ref{fieldzero})), using
unitary operators of Eqs. (\ref{u_two} ) and (\ref{u_one}), state of
composite system after interaction time $t=\tau /g_{c}$ reads as%
\begin{eqnarray}
\widehat{\rho }(\tau ) &=&\widehat{U}_{1}(\tau )\otimes \widehat{U}_{2}(\tau
)\widehat{\rho }_{A_{1}A_{2}B}(0)\otimes \widehat{\rho }_{_{F}}(0)\widehat{U}%
_{1}^{\dag }(\tau )\otimes \widehat{U}_{2}^{\dag }(\tau )  \notag \\
&=&\left( \frac{1}{\cosh s}\right) ^{2}\overset{\infty }{\underset{n,m=0}{%
\sum }}\underset{k,l=0}{\overset{\min \left[ n,m\right] }{\;\sum }}(\tanh
s)^{n+m}G_{kl}^{nm}(\theta )\left\vert \Phi _{A_{1}A_{2}B}^{n-k,n-l}(\tau
)\right\rangle \left\langle \Phi _{A_{1}A_{2}B}^{m-k,m-l}(\tau )\right\vert ,
\label{rotau}
\end{eqnarray}%
where%
\begin{equation}
\left\vert \Phi _{A_{1}A_{2}B}^{n-k,n-l}(\tau )\right\rangle =\hat{U}%
_{12}^{n-k,n-l}(\tau )\left\vert \Phi _{A_{1}A_{2}}(0),n-k\right\rangle
_{c_{1}}\left\vert \Phi _{B}(0),n-l\right\rangle _{c_{2}}.
\end{equation}

The information about the effective evolution of three atom entanglement is
contained in the state operator $\widehat{\rho }_{A_{1}A_{2}B}(\tau )$,
obtained from $\widehat{\rho }(\tau )$ upon partial trace over the CV
degrees of freedom that is

\begin{equation}
\widehat{\rho }_{A_{1}A_{2}B}(\tau )=Tr_{F}(\widehat{\rho }(\tau )).
\end{equation}%
The matrix $\widehat{\rho }_{A_{1}A_{2}B}(\tau )$ is used to analyze,
numerically, the entanglement generation between the remote qubit $B$ and
the pair of qubits $A_{1}A_{2}$.

\subsection{ Atoms in Initial State $\left\vert \Phi _{1}^{\protect\alpha %
}(0)\right\rangle =\protect\sqrt{\protect\alpha }\left\vert 000\right\rangle
+\protect\sqrt{\left( 1-\protect\alpha \right) }\left\vert 110\right\rangle $%
}

We consider two different types of atomic initial states. Firstly, two atoms
in cavity $c_{1}$ are prepared in state%
\begin{equation}
\left\vert \Phi _{A_{1}A_{2}}^{\alpha }(0)\right\rangle =\sqrt{\alpha }%
\left\vert 2,-2\right\rangle +\sqrt{\left( 1-\alpha \right) }\left\vert
2,2\right\rangle ,
\end{equation}%
while the atom in cavity $c_{2}$ is in ground state at $t=0$. The two atoms
have varying degree of entanglement for $0<\alpha <1$.\ States $\left\vert
\Phi _{A_{1}A_{2}}^{0}(0)\right\rangle $ and $\left\vert \Phi
_{A_{1}A_{2}}^{1}(0)\right\rangle $ are separable states. Interaction of
atoms with two mode squeezed field is known to generate entanglement of
qubit $B$ with pair $A_{1}A_{2}$ \cite{reis09}, however, the role of initial
entanglement of atoms in entanglement generation is not known. Our object is
to investigate if initial entanglement of atoms serves as a catalyst in the
process of entanglement transfer or hinders it. The maximal value of
entanglement generated and inhibition of zero entanglement zones are
considered as the pointers or indicators of such effects.

Associating computational basis state $\left\vert 0\right\rangle $ to atomic
ground state and $\left\vert 1\right\rangle $ to an atom in excited state,
the initial three atom state is 
\begin{eqnarray*}
\left\vert \Phi _{1}^{\alpha }(0)\right\rangle  &=&\left\vert \Phi
_{A_{1}A_{2}}^{\alpha }(0)\right\rangle \left\vert \Phi _{B}(0)\right\rangle
=\sqrt{\alpha }\left\vert 000\right\rangle +\sqrt{\left( 1-\alpha \right) }%
\left\vert 110\right\rangle , \\
\widehat{\rho }_{A}^{\alpha }(0) &=&\left\vert \Phi _{1}^{\alpha
}(0)\right\rangle \left\langle \Phi _{1}^{\alpha }(0)\right\vert .
\end{eqnarray*}%
The state of atom-field composite system after interaction time $t$ ,
obtained by using Eq. (\ref{rotau}), reads as%
\begin{equation}
\widehat{\rho }^{\alpha }(\tau )=\left( \frac{1}{\cosh s}\right) ^{2}\overset%
{\infty }{\underset{n,m=0}{\sum }}\underset{k,l=0}{\overset{\min \left[ n,m%
\right] }{\;\sum }}(\tanh s)^{n+m}G_{kl}^{nm}(\theta )\left\vert \Phi
_{A_{1}A_{2}B}^{n-k,n-l}(\tau )\right\rangle _{\alpha }\left\langle \Phi
_{A_{1}A_{2}B}^{m-k,m-l}(\tau )\right\vert .  \label{ro1tau}
\end{equation}%
Exact analytic expression for $\left\vert \Phi _{A_{1}A_{2}B}^{n-k,n-l}(\tau
)\right\rangle _{\alpha }$\ is given in Appendix (\ref{A1}). The state
operator for atomic qubits $A_{1}$, $A_{2}$, and $B$ is obtained from $%
\widehat{\rho }^{\alpha }(\tau )$ by tracing out the field modes that is%
\begin{equation}
\widehat{\rho }_{A}^{\alpha }(\tau )=Tr_{F}(\widehat{\rho }^{\alpha }(\tau
)).
\end{equation}%
The matrix $\rho _{A}^{\alpha }(\tau )$ in the basis $\left\vert
2,-2\right\rangle _{1}\left\vert 1,-1\right\rangle _{2}$, $\left\vert
2,0\right\rangle _{1}\left\vert 1,-1\right\rangle _{2}$, $\left\vert
2,2\right\rangle _{1}\left\vert 1,-1\right\rangle _{2}$, $\left\vert
2,-2\right\rangle _{1}\left\vert 1,1\right\rangle _{2}$, $\ \left\vert
2,0\right\rangle _{1}\left\vert 1,1\right\rangle _{2}$, $\ \left\vert
2,2\right\rangle _{1}\left\vert 1\right\rangle _{2}$, has the form%
\begin{equation}
\rho _{A}^{\alpha }(\tau )=\left( 
\begin{array}{cccccc}
\left( \rho _{A}^{\alpha }(\tau )\right) _{11} & 0 & \left( \rho
_{A}^{\alpha }(\tau )\right) _{13} & 0 & \left( \rho _{A}^{\alpha }(\tau
)\right) _{15} & 0 \\ 
0 & \left( \rho _{A}^{\alpha }(\tau )\right) _{22} & 0 & \left( \rho
_{A}^{\alpha }(\tau )\right) _{42} & 0 & \left( \rho _{A}^{\alpha }(\tau
)\right) _{26} \\ 
\left( \rho _{A}^{\alpha }(\tau )\right) _{13} & 0 & \left( \rho
_{A}^{\alpha }(\tau )\right) _{33} & 0 & \left( \rho _{A}^{\alpha }(\tau
)\right) _{53} & 0 \\ 
0 & \left( \rho _{A}^{\alpha }(\tau )\right) _{42} & 0 & \left( \rho
_{A}^{\alpha }(\tau )\right) _{44} & 0 & \left( \rho _{A}^{\alpha }(\tau
)\right) _{46} \\ 
\left( \rho _{A}^{\alpha }(\tau )\right) _{15} & 0 & \left( \rho
_{A}^{\alpha }(\tau )\right) _{53} & 0 & \left( \rho _{A}^{\alpha }(\tau
)\right) _{55} & 0 \\ 
0 & \left( \rho _{A}^{\alpha }(\tau )\right) _{26} & 0 & \left( \rho
_{A}^{\alpha }(\tau )\right) _{46} & 0 & \left( \rho _{A}^{\alpha }(\tau
)\right) _{66}%
\end{array}%
\right) .  \label{rho1}
\end{equation}

\begin{figure}[t]
\centering \includegraphics[width=6in,height=5in]{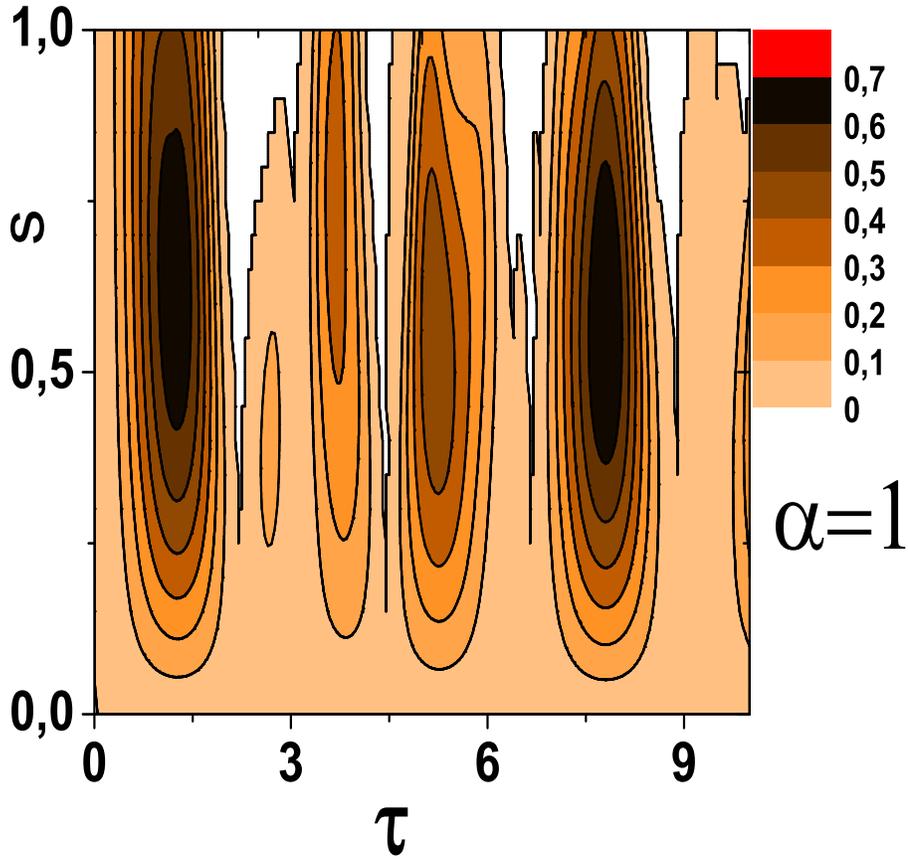} 
\caption{Contour plot of global negativity $N_{G}^{B}(\protect\rho %
_{A}^{alpha=1}\left( \protect\tau \right) )$ as a function of $s$ and $%
\protect\tau $. Area in white represents zero negativity.}
\label{fig1}
\end{figure}

\begin{figure}[t]
\centering \includegraphics[width=6in,height=5in,angle=0]{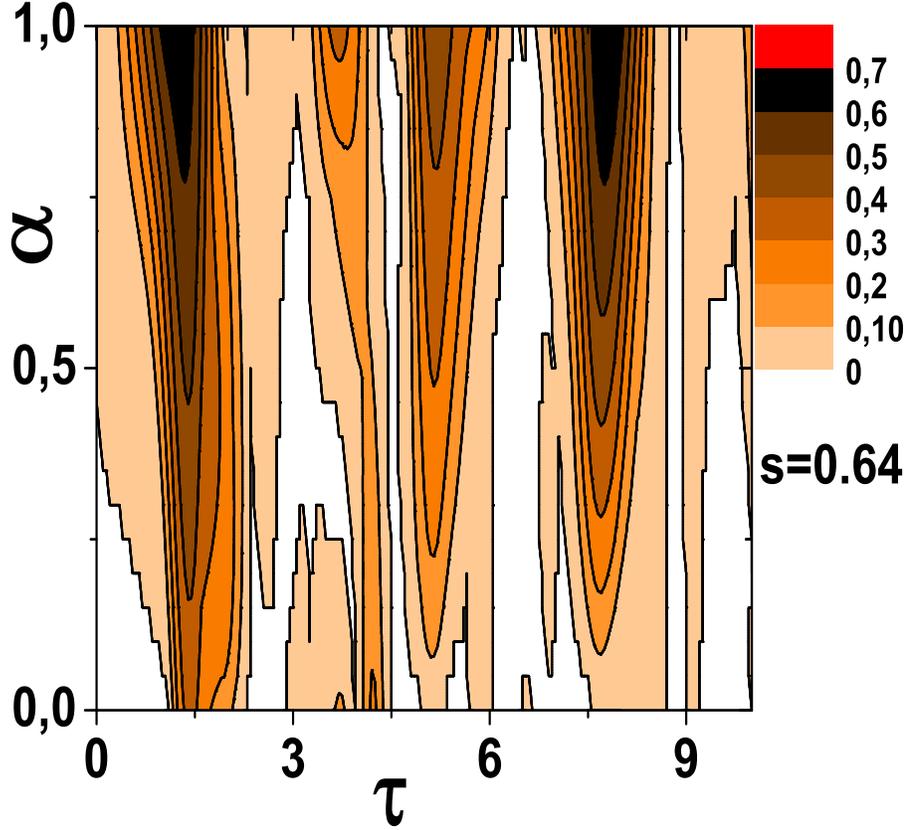}
\caption{Contour plot of global negativity $N_{G}^{B}(\protect\rho_{A}^{%
\protect\alpha}\left( \protect\tau \right) )$ versus $\protect\alpha $ and $%
\protect\tau $ for $s=0.64$.}
\label{fig2}
\end{figure}

\subsection{Atoms in Initial State $\left\vert \Phi _{2}(0)\right\rangle =%
\frac{1}{\protect\sqrt{2}}\left( \left\vert 010\right\rangle +\left\vert
100\right\rangle \right) $}

Another possibility, where pair of qubits A$_{1}$ and A$_{2}$ are in an
entangled state at $t=0$ arises with the atoms prepared initially in states $%
\left\vert \Phi _{A_{1}A_{2}}(0)\right\rangle =\left\vert 2,0\right\rangle ,$
and $\left\vert \Phi _{B}(0)\right\rangle =\left\vert 1,-1\right\rangle ,$%
that is 
\begin{equation*}
\left\vert \Phi _{2}(0)\right\rangle =\frac{1}{\sqrt{2}}\left( \left\vert
010\right\rangle +\left\vert 100\right\rangle \right) ,\quad \widehat{\rho }%
_{A}^{II}(0)=\left\vert \Phi _{2}(0)\right\rangle \left\langle \Phi
_{2}(0)\right\vert .
\end{equation*}%
Using Eq. (\ref{rotau} ), the state of composite system after interaction
time $t$ is found to be%
\begin{eqnarray}
\widehat{\rho }_{A}^{II}(\tau ) &=&\hat{U}_{12}(\tau )\widehat{\rho }%
_{A}^{II}(0)\otimes \widehat{\rho }_{_{F}}(0)\hat{U}_{12}^{\dagger }(\tau ) 
\notag \\
&=&\left( \frac{1}{\cosh s}\right) ^{2}\overset{\infty }{\underset{n,m=0}{%
\sum }}\underset{k,l=0}{\overset{\min \left[ n,m\right] }{\;\sum }}(\tanh
s)^{n+m}G_{kl}^{nm}(\theta )\left\vert \Phi _{A_{1}A_{2}B}^{n-k,n-l}(\tau
)\right\rangle \left\langle \Phi _{A_{1}A_{2}B}^{m-k,m-l}(\tau )\right\vert ,
\label{ro2tau}
\end{eqnarray}%
with $\left\vert \Phi _{A_{1}A_{2}B}^{n-k,n-l}(\tau )\right\rangle $ as
listed in Appendix (\ref{A2}). The corresponding atomic density operator $%
\widehat{\rho }_{A}^{II}(\tau )$, in the basis $\left\vert 2,-2\right\rangle
_{c_{1}}\left\vert 1,-1\right\rangle _{c_{2}}$, $\left\vert 2,0\right\rangle
_{c_{1}}\left\vert 1,-1\right\rangle _{c_{2}}$, $\left\vert 2,2\right\rangle
_{c_{1}}\left\vert 1,-1\right\rangle _{c_{2}}$, $\left\vert
2,-2\right\rangle _{c_{1}}\left\vert 1,1\right\rangle _{c_{2}}$, $\left\vert
2,0\right\rangle _{c_{1}}\left\vert 1,1\right\rangle _{c_{2}}$, $\left\vert
2,2\right\rangle _{c_{1}}\left\vert 1,1\right\rangle _{c_{2}}$, reads as%
\begin{equation}
\rho _{A}^{II}(\tau )=\left( 
\begin{array}{cccccc}
\left( \rho _{A}^{II}(\tau )\right) _{11} & 0 & 0 & 0 & \left( \rho
_{A}^{II}(\tau )\right) _{15} & 0 \\ 
0 & \left( \rho _{A}^{II}(\tau )\right) _{22} & 0 & 0 & 0 & \left( \rho
_{A}^{II}(\tau )\right) _{26} \\ 
0 & 0 & \left( \rho _{A}^{II}(\tau )\right) _{33} & 0 & 0 & 0 \\ 
0 & 0 & 0 & \left( \rho _{A}^{II}(\tau )\right) _{44} & 0 & 0 \\ 
\left( \rho _{A}^{II}(\tau )\right) _{15} & 0 & 0 & 0 & \left( \rho
_{A}^{II}(\tau )\right) _{55} & 0 \\ 
0 & \left( \rho _{A}^{II}(\tau )\right) _{26} & 0 & 0 & 0 & \left( \rho
_{A}^{II}(\tau )\right) _{66}%
\end{array}%
\right) .  \label{rho2}
\end{equation}

\begin{figure}[t]
\centering %
\includegraphics[width=6in,height=5in,angle=0]{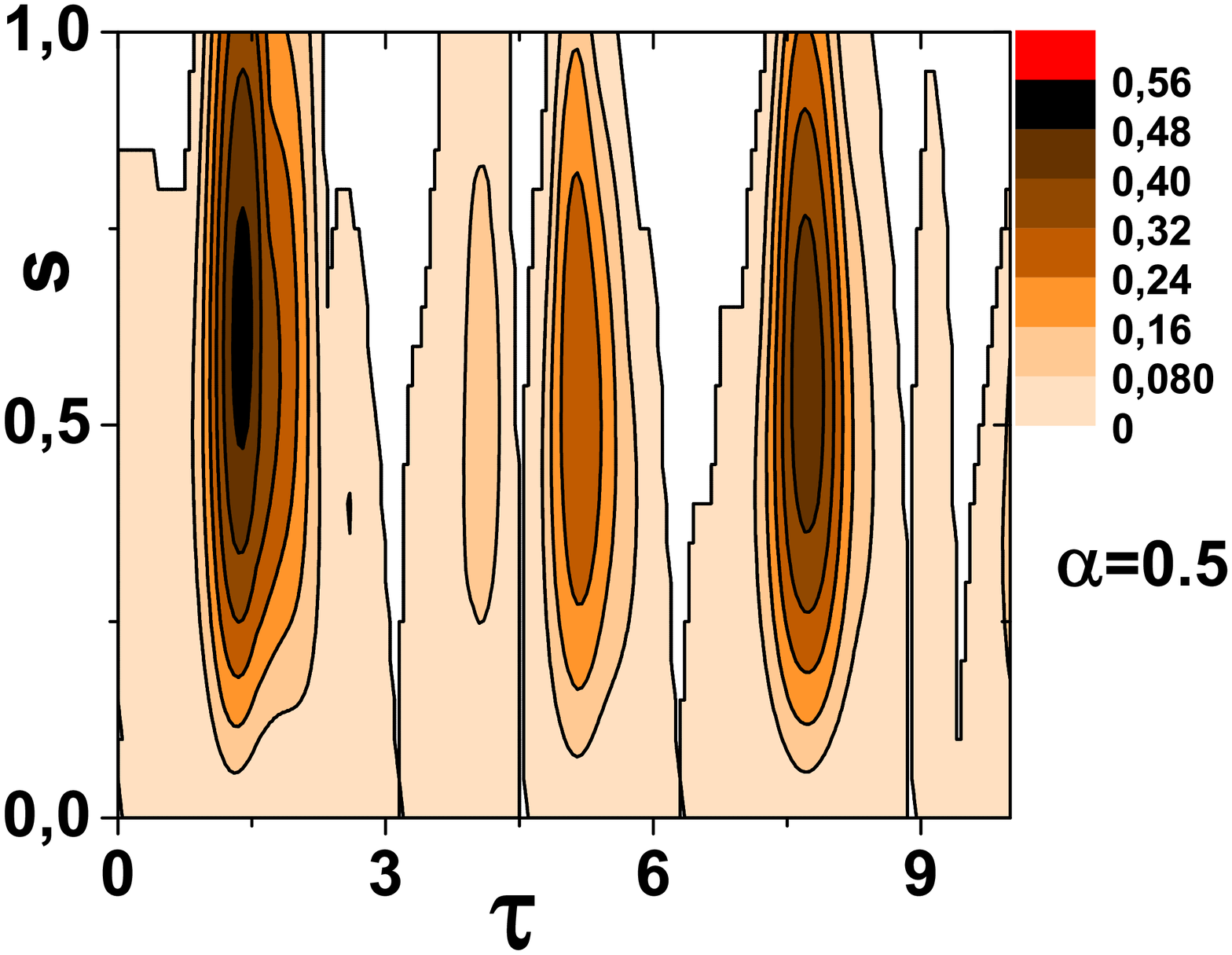}
\caption{Contour plot of global negativity $N_{G}^{B}(\protect\rho _{A}^{%
\protect\alpha=0.5}\left( \protect\tau \right) )$ as a function of $s$ and $%
\protect\tau $.}
\label{fig3}
\end{figure}

\begin{figure}[t]
\centering \includegraphics[width=6in,height=5in,angle=0]{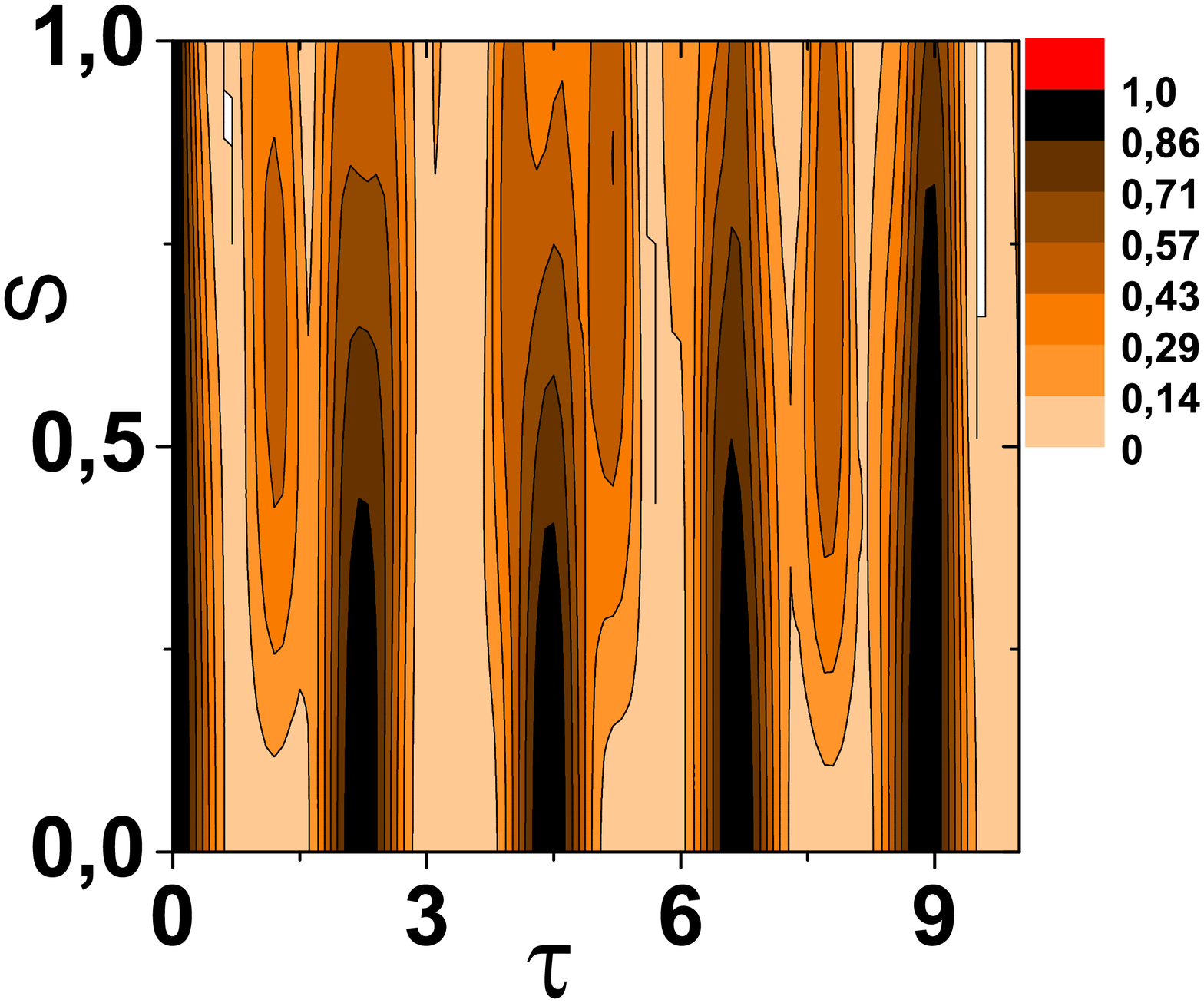}
\caption{Contour plot of global negativity $N_{G}^{A_{1}}(\protect\rho %
^{II}\left( \protect\tau \right) )$ as a function of $s$ and $\protect\tau $%
. }
\label{fig4}
\end{figure}

\section{Three qubit entanglement dynamics}

In our earlier article \cite{reis09}, the three atom system in separable
initial state was found to develop $W-$like entanglement exhibiting
entanglement sudden death (ESD) and revival (ESR). Entanglement sudden death
refers to disappearance of entanglement of the state in finite time and was
reported by Yu and Eberly \cite{yu04,yu06} for the first time. ESD has been
observed experimentally for entangled photon pairs \cite{alme07}, and atomic
ensembles \cite{laur07}. It is a hindrance to using the system for
implementing useful protocols. We analyze, numerically, the dependence of
three qubit entanglement generation on the degree of squeezing of
electromagnetic field and initial state entanglement of qubits A$_{1}$ and A$%
_{2}$. Since the focus is on the effect of initial state entanglement of
qubits in cavity $c_{1}$ on entanglement transfer from CV field and
generation of tripartite distributed entanglement, for simplicity, we
neglect the effect of the mirror transmittance, i.e., consider $T(\theta )=1$%
. For appropriate choice of squeeze parameter the time interval between ESD
and ESR (Entanglement sudden revival) is found to become shorter or
disappear altogether.

\subsection{Entanglement of qubit $B$ with pair $A_{1}A_{2}$}

Fortran codes were written to calculate, numerically, the system dynamics
for initial states with varying degree of field state entanglement and two
atom entanglement in cavity c$_{2}$, using analytic expressions obtained in
section II. To analyze the entanglement of remote qubit $B$ in cavity c$_{2}$
with qubits $A_{1}A_{2}$ in cavity c$_{1}$, the global negativity of
partially transposed state operator was calculated. Writing a general three
qubit as%
\begin{equation}
\widehat{\rho }=\sum_{\substack{ i_{1}i_{2}i_{3} \\ j_{1}j_{2}j_{3}}}%
\left\langle i_{1}i_{2}i_{3}\right\vert \widehat{\rho }\left\vert
j_{1}j_{2}j_{3}\right\rangle \left\vert i_{1}i_{2}i_{3}\right\rangle
\left\langle j_{1}j_{2}j_{3}\right\vert ,  \label{rho}
\end{equation}%
where $\left\vert i_{1}i_{2}i_{3}\right\rangle $ are the basis vectors
spanning $2^{3}$ dimensional Hilbert space, the global partial transpose
with respect to qubit $B$ is constructed from the matrix elements of $%
\widehat{\rho }$ through 
\begin{equation}
\left\langle i_{1}i_{2}i_{3}\right\vert \widehat{\rho }_{G}^{T_{B}}\left%
\vert j_{1}j_{2}j_{3}\right\rangle =\left\langle i_{1}i_{2}j_{3}\right\vert 
\widehat{\rho }\left\vert j_{1}j_{2}i_{3}\right\rangle .
\end{equation}%
The state of qubit at location $A_{m}$ is labelled by $i_{m}=0$ and $1,$
where $m=1,2$. The state of qubit $B$ is labelled by $i_{3}=0$,$1$. \ Global
negativity is defined as 
\begin{equation}
N_{G}^{A_{p}}=\left( \left\Vert \widehat{\rho }_{G}^{T_{A_{p}}}\right\Vert
_{1}-1\right) ,
\end{equation}%
where $\left\Vert \widehat{\rho }\right\Vert _{1}$ is the trace norm of $%
\widehat{\rho }$. Global negativity lies in the range $0$ for a separable
state to $1$ for a maximally entangled state. The negativity \cite%
{zycz98,eise99,vida02} of $\widehat{\rho }_{G}^{T_{B}}$, based on
Peres-Horodecki criterion \cite{pere96,horo98} is a natural entanglement
measure and has been shown to be an entanglement monotone \cite{vida02}. As
the qubits $A_{1}$ and $A_{2}$ are always in a symmetric initial state, the
entanglement of qubit B with either of the qubits is the same. Therefore a
negative partial transpose of $\widehat{\rho }_{A}(\tau )$ with respect to
qubit $B$ indicates tripartite entanglement. A three qubit state may have
GHZ-like or W-like tripartite \cite{dur00} entanglement. In ref \cite{shar07}
it has been shown that a global partial transpose may be written in terms of
a two way partial transpose, a three-way partial transpose and the state
operator.We notice that for states $\widehat{\rho }_{A}^{\alpha }(\tau )$
and $\widehat{\rho }_{A}^{II}(\tau )$ partial transposition involves, only,
matrix elements $\left\langle i_{1}i_{2}j_{3}\right\vert \widehat{\rho }%
_{A}(\tau )\left\vert j_{1}j_{2}i_{3}\right\rangle $ with $%
K=\sum\limits_{i,j=1,i<j}^{3}\left( 1-\delta _{i,j}\right) =2$. In other
words the global partial transpose $\left( \rho _{A}(\tau )\right)
_{G}^{T_{B}}$ is equal to two way partial transpose $\left( \rho _{A}(\tau
)\right) _{2}^{T_{B}}$ \cite{shar07}. Therefore the entanglement of mixed
state is similar to W-like entanglement of three qubit pure states. It is a
natural consequence of the fact that no direct three atom interaction takes
place.

Firstly, we discuss the numerical results of global negativity of partial
transpose of matrix $\rho _{A}^{\alpha }(\tau )$ with respect to remote
qubit $B$. Fig. (\ref{fig1}) is a contour plot of $N_{G}^{B}(\rho
_{A}^{\alpha =1}\left( \tau \right) )$ as a function of squeeze parameter $s$
and interaction parameter $\tau $, for two atoms in separable state $%
\left\vert \Phi _{1}^{\alpha =1.0}(0)\right\rangle $. Entanglement of remote
qubit at peak value is found to increase with $s$, being optimum for $s=0.64$%
, where peak value of $N_{G}^{B}(\rho _{a}^{\alpha }\left( \tau \right) )$
is around $0.7$ \ However, for a fixed value of $s$ the regions with
continuously zero negativity (white) alternate with regions having finite
global negativity and the meeting points represent values of $\tau $ for
which entanglement sudden death or revival of entanglement occurs. A contour
plot of $N_{G}^{B}(\rho _{A}^{\alpha }\left( \tau \right) )$ as a function
of $\alpha $, $\tau $ and $s=0.64$ displayed in Fig. (\ref{fig2}) reveals
that for $\alpha <1$ that is an entangled state $\left\vert \Phi
_{A_{1}A_{2}}^{\alpha }(0)\right\rangle ,$ the global negativity strongly
depends on value of $\alpha $. The peak value of $N_{G}^{B}(\rho
_{A}^{\alpha }\left( \tau \right) )$ is found to decrease as $\alpha
\rightarrow 0$ and regions with zero entanglement become wider in comparison
with that for $\alpha =1$. A comparison of linear entropy of $\rho
_{A}^{\alpha =1}\left( \tau \right) $ and $\rho _{A}^{\alpha \neq 1}\left(
\tau \right) $ at peak value of $N_{G}^{B}(\rho _{A}^{\alpha }\left( \tau
\right) )$ shows a larger value for $\rho _{A}^{\alpha \neq 1}\left( \tau
\right) $, indicating that initial entanglement generates a noisier three
qubit state than the state $\rho _{A}^{\alpha =1}\left( \tau \right) $.

Figure \ref{fig3} displays the negativity $N_{G}^{B}(\rho _{A}^{\alpha
=0.5}\left( \tau \right) )$, versus compression parameter $s$ and
interaction parameter $\tau $, for the initial state $\left\vert \Phi
_{1}^{\alpha =0.5}(0)\right\rangle $. In this case maximum value of $%
N_{G}^{B}(\rho _{A}^{\alpha =0.5}\left( \tau \right) )=0.5$ corresponds to $%
s=0.64$ and occurs at $\tau =7.75$. The interaction time after which
entanglement dies is $\tau =2.45$, for $s=0.64$. When we use a compression
parameter smaller than $0.64$ zero negativity regions shrink pointing to a
decrease in noise however the peak value of negativity tends to zero as
well. 
\begin{figure}[t]
\centering \includegraphics[width=6in,height=5in,angle=0]{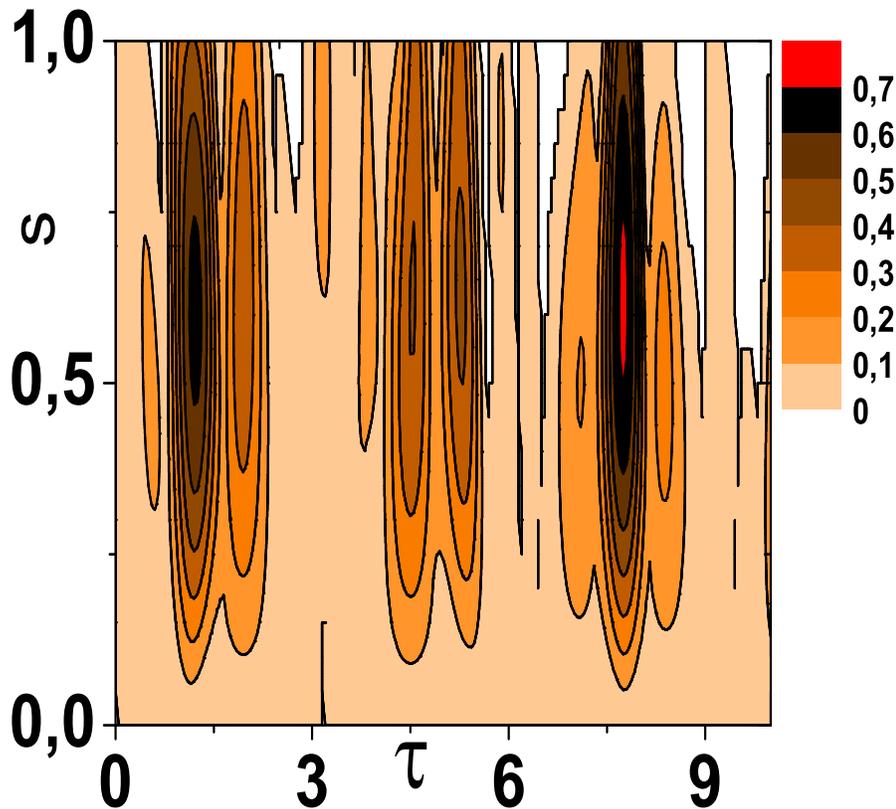}
\caption{Contour plot of global negativity $N_{G}^{B}(\protect\rho %
^{II}\left( \protect\tau \right) )$ as a function of $s$ and $\protect\tau $%
. }
\label{fig5}
\end{figure}

\subsection{Entanglement of qubit $A_{1}$ with $A_{2}B$}

Fig. \ref{fig4} is a contour plot of global negativity $N_{G}^{A_{1}}(\rho
_{A}^{II}\left( \tau \right) )$ as a function of squeeze parameter $s$ and
interaction parameter $\tau $, for initial state $\left\vert \Phi
_{2}(0)\right\rangle $. As expected the interaction with squeezed field
results in a decoherence of initial entanglement of qubit $A_{1}$.
Decoherence becomes more pronounced with increment in the value of parameter 
$s$. For $s\geq 0.5$, the peak value of $N_{G}^{A_{1}}$ is always less than
one for the values of $\tau $ in the range shown in the plot. Fig. \ref{fig5}
displays $N_{G}^{B}(\rho _{A}^{II}(\tau ))$ as a function of squeeze
parameter $s$ and interaction parameter $\tau $, for initial state $%
\left\vert \Phi _{2}(0)\right\rangle $. While part of the initial state
entanglement of qubit $A_{1}$ is transferred to field degrees a small part
goes to increase the entanglement of remote qubit $B$ with pair of qubits in
cavity one. In this case the maximum value of negativity is $0.73$, which
corresponds to $s=0.64$ and $\tau =7.75$.

\begin{figure}[t]
\centering \includegraphics[width=6in,height=5in,angle=0]{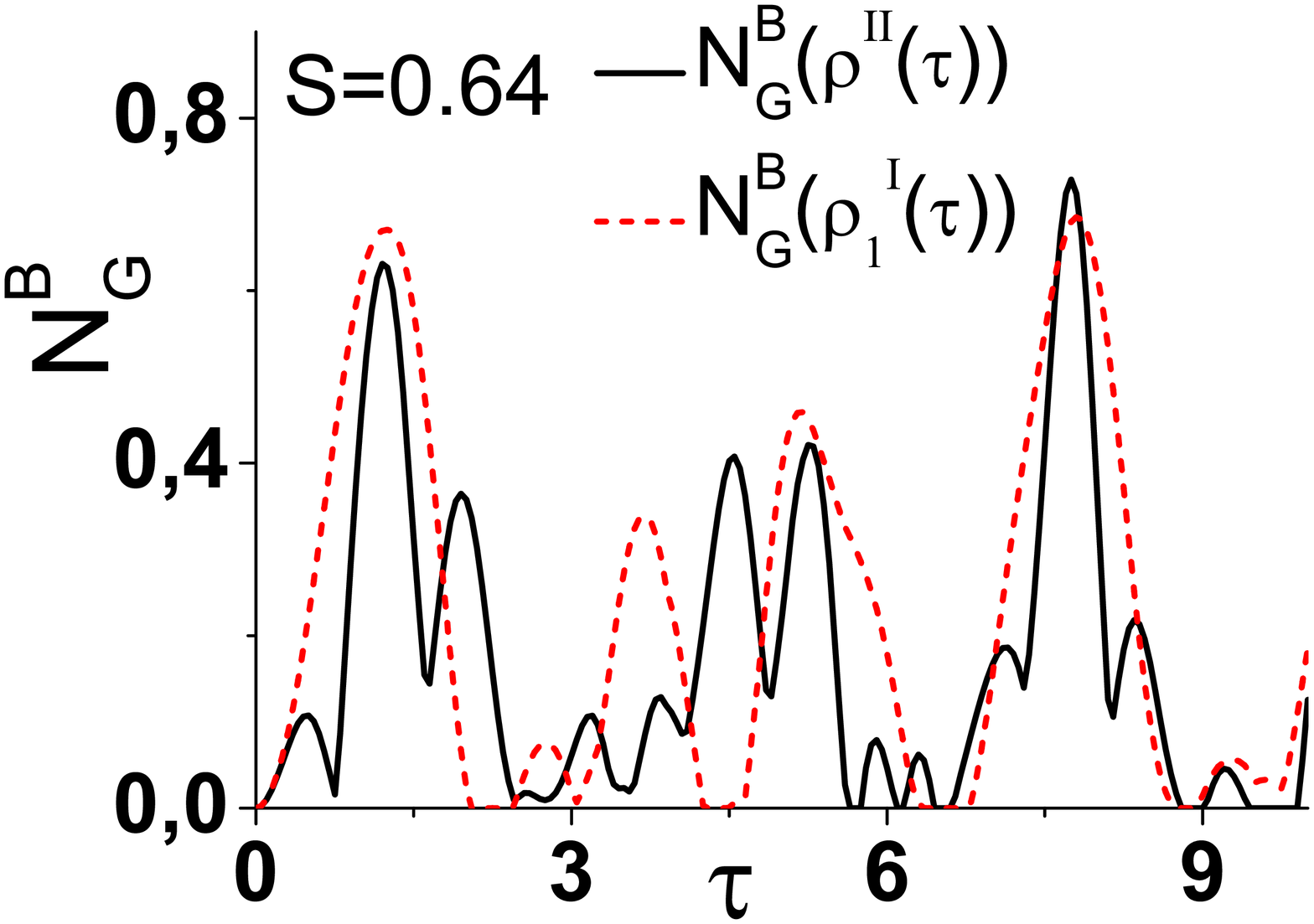} 
\caption{$N_{G}^{B}$ versus $\protect\tau $ \ for $s=0.64$, for initial
states $\left\vert \Phi _{2}\right\rangle $(solid line) and $\left\vert \Phi
_{1}^{\protect\alpha =1.0}\right\rangle $ (dashed line). }
\label{fig6}
\end{figure}

\begin{figure}[t]
\centering \includegraphics[width=6in,height=5in,angle=0]{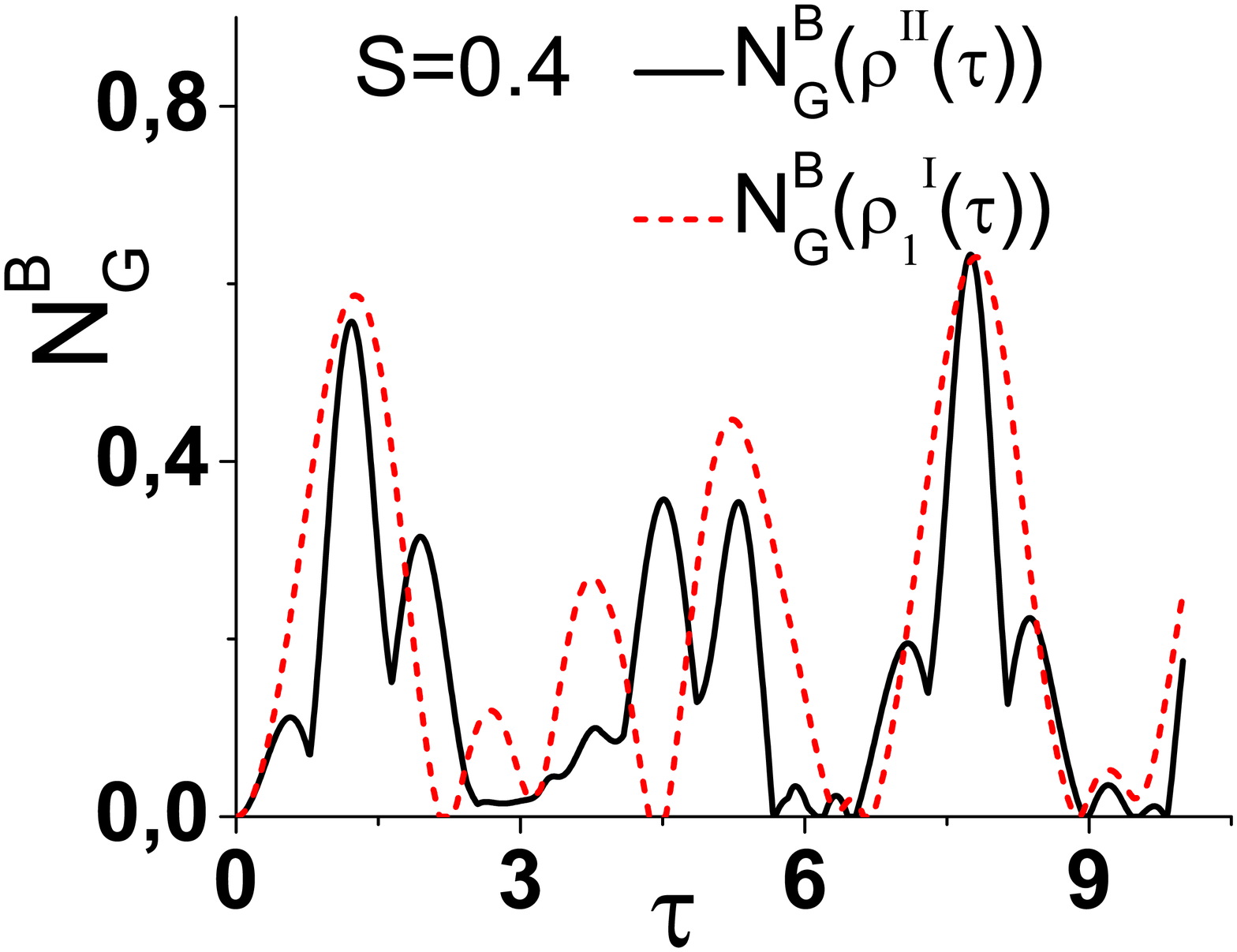} 
\caption{$N_{G}^{B}$ versus $\protect\tau $ \ for $s=0.4$, for initial
states $\left\vert \Phi _{2}\right\rangle $(solid line) and $\left\vert \Phi
_{1}^{\protect\alpha =1.0} \right\rangle $ (dashed line). }
\label{fig7}
\end{figure}

Figs. (\ref{fig6}) and (\ref{fig7}) display $N_{G}^{B}(\rho _{A}^{\alpha
=1}\left( \tau \right) )$ and $N_{G}^{B}(\rho _{A}^{II}(\tau ))$ plotted for 
$s=0.64$ and $s=0.4$, respectively. It is seen that for initial state $%
\left\vert \Phi _{2}(0)\right\rangle $ the interaction time after which ESD
occurs is longer than that for the case of separable initial state. The peak
value of $N_{G}^{B}(\rho _{A}^{II}(\tau ))=0.73$ is comparable to peak value
of $N_{G}^{B}(\rho _{A}^{\alpha =1}\left( \tau \right) )=0.7$ when $s=0.64$.
Figure \ref{fig7} shows a decreased peak value of around $0.63$ both for $%
N_{G}^{B}(\rho _{A}^{\alpha =1}\left( \tau \right) )$ and $N_{G}^{B}(\rho
_{A}^{II}(\tau ))$. In the case of initial state $\left\vert \Phi
_{2}(0)\right\rangle $, lowering the value of $s$ increases the value of
interaction parameter $\tau $ for which ESD occurs that is an increase in
three qubit quantum correlations. The initially entangled state $\left\vert
\Phi _{2}(0)\right\rangle $ with $s\leq 0.64$ is better suited to generate
entanglement between the remote qubit $B$ and the pair of qubits in cavity c$%
_{1}$ in comparison with initial states $\left\vert \Phi _{1}^{\alpha
}(0)\right\rangle $. 

\section{Conclusions}

In this article, we have examined the entanglement generation\ due to
interaction of an entangled pair\ of two level atoms $A_{1}A_{2}$ in cavity $%
c_{1}$ and an atom $B$ in a remote cavity $c_{2}$ with two mode squeezed
field shared by the cavities. Three qubit mixed state entanglement dynamics
is a function of squeeze parameter value and initial state entanglement of
the qubit pair $A_{1}A_{2}$. Firstly, starting from different initial
states, analytical expressions have been obtained for three qubit mixed
state. The entanglement of qubits $A_{1}A_{2}B$ is that of a W-like state,
no genuine tripartite entanglement being generated. Numerical values of
global negativity of partial transpose with respect to qubit $B$ are used to
search for initial states and squeeze parameter values that generate highly
entangled states. We notice that the two mode squeezed field shared by two
cavities generates entanglement of qubit $B$ with pair of qubits $A_{1}A_{2}$
which reaches a peak value followed by sudden disappearance and revival. The
dynamics of entanglement generation, sudden death and revival strongly
depends on the three qubit initial state.

When both the atoms in cavity $c_{1}$ are prepared in ground state at $t=0$,
the entanglement of remote qubit is found to increase with $s$, being
optimum for $s=0.64$. This is significant from the point of view of
practical utility of the mixed state entanglement. Starting with the pair of
atoms $A_{1}A_{2}$ prepared initially in an entangled state $\left\vert \Phi
_{1}^{\alpha }(0)\right\rangle =\sqrt{\alpha }\left\vert 00\right\rangle +%
\sqrt{1-\alpha }\left\vert 11\right\rangle $, the peak value of entanglement
between the remote qubit $B$ and pair of qubits $A_{1}A_{2}$ decreases as $%
\alpha \rightarrow 0$. With atoms $A_{1}A_{2}$ in state $\left\vert \Phi
_{2}(0)\right\rangle =\left( \left\vert 10\right\rangle +\left\vert
01\right\rangle \right) /\sqrt{2}$ in first cavity, atom field interaction\
generates a three qubit mixed state with atom $B$ in second cavity highly
entangled to Bell pair $A_{1}A_{2}$. In this case, entanglement transfer
from the pair $A_{1}A_{2}$ partially compensates for the loss of
entanglement due to state reduction. As a result, three qubit entanglement
dynamics displays, relatively, longer sudden death free intervals. For a
given value of squeeze parameter, the entanglement of remote qubit $B$ with
qubit pair $A_{1}A_{2}$ is of the same order as in the case when all three
qubits are in a separable state, initially. However, a smaller value $s=0.4$
under similar conditions generates a three qubit mixed state with comparable
entanglement but lesser noise. We conclude that initially entangled state $%
\left\vert \Phi _{2}(0)\right\rangle $ with $s\leq 0.64$ is better suited to
generate entanglement between the remote qubit $B$ and the pair of qubits $%
A_{1}A_{2}$ in cavity $c_{1}$ in comparison with initial states $\left\vert
\Phi _{1}^{\alpha }(0)\right\rangle $.

Financial support from, Capes Brazil, CNPq Brazil, Faep Uel Brazil, and Funda%
\c{c}\~{a}o Araucaria Pr Brazil is acknowledged.

\appendix

\label{A1}

Analytic expression for $\left\vert \Phi _{A_{1}A_{2}B}^{n-k,n-l}(\tau
)\right\rangle _{\alpha }$ in Eq (\ref{ro1tau}) reads as%
\begin{eqnarray}
&&\left\vert \Phi _{A_{1}A_{2}B}^{n-k,n-l}(\tau )\right\rangle _{\alpha } 
\notag \\
&=&\hat{U}_{12}^{n-k,n-l}(\tau )\left\vert \Phi _{A_{1}A_{2}}^{\alpha
}(0)\right\rangle \left\vert 1,-1\right\rangle \left\vert
n-k,n-l\right\rangle  \notag \\
&=&\sqrt{\alpha }\cos \left( C_{nl}\tau \right) \frac{\left[ B_{nk}^{2}\cos
\left( f_{nk}\tau \right) +A_{nk}^{2}\right] }{A_{nk}^{2}+B_{nk}^{2}}%
\left\vert 2,-2\right\rangle \left\vert 1,-1\right\rangle \left\vert
n-k,n-l\right\rangle  \notag \\
&&-i\sqrt{\alpha }\left( \cos C_{nl}\tau \right) B_{nk}\frac{\sin \left(
f_{nk}\tau \right) }{\sqrt{\left( A_{nk}^{2}+B_{nk}^{2}\right) }}\left\vert
2,0\right\rangle \left\vert 1,-1\right\rangle \left\vert
n-k-1,n-l\right\rangle  \notag \\
&&+\sqrt{\alpha }\left( \cos C_{nl}\tau \right) A_{nk}B_{nk}\frac{\left[
\cos \left( f_{nk}\tau \right) -1\right] }{A_{nk}^{2}+B_{nk}^{2}}\left\vert
2,2\right\rangle \left\vert 1,-1\right\rangle \left\vert
n-k-2,n-l\right\rangle  \notag \\
&&-i\sqrt{\alpha }\left( \sin C_{nl}\tau \right) \frac{\left[ B_{nk}^{2}\cos
\left( f_{nk}\tau \right) +A_{nk}^{2}\right] }{A_{nk}^{2}+B_{nk}^{2}}%
\left\vert 2,-2\right\rangle \left\vert 1,1\right\rangle \left\vert
n-k,n-l-1\right\rangle  \notag \\
&&-\sqrt{\alpha }\left( \sin C_{nl}\tau \right) B_{nk}^{2}\frac{\sin \left(
f_{nk}\tau \right) }{\sqrt{\left( A_{nk}^{2}+B_{nk}^{2}\right) }}\left\vert
2,0\right\rangle \left\vert 1,1\right\rangle \left\vert
n-k-1,n-l-1\right\rangle  \notag \\
&&-i\sqrt{\alpha }\left( \sin C_{nl}\tau \right) A_{nk}B_{nk}\frac{\left[
\cos \left( f_{nk}\tau \right) -1\right] }{A_{nk}^{2}+B_{nk}^{2}}\left\vert
2,2\right\rangle \left\vert 1,1\right\rangle \left\vert
n-k-2,n-l-1\right\rangle  \notag \\
&&+\sqrt{1-\alpha }\left( \cos C_{nl}\tau \right) A_{n+2k}B_{n+2k}\frac{%
\left[ \cos \left( f_{n+2k}\tau \right) -1\right] }{A_{n+2k}^{2}+B_{n+2k}^{2}%
}\left\vert 2,-2\right\rangle \left\vert 1,-1\right\rangle \left\vert
n-k+2,n-l\right\rangle  \notag \\
&&-i\sqrt{\left( 1-\alpha \right) }\left( \cos C_{nl}\tau \right) A_{n+2k}%
\frac{\sin \left( f_{n+2k}\tau \right) }{\sqrt{\left(
A_{n+2k}^{2}+B_{n+2k}^{2}\right) }}\left\vert 2,0\right\rangle \left\vert
1,-1\right\rangle \left\vert n-k+1,n-l\right\rangle  \notag \\
&&+\sqrt{\left( 1-\alpha \right) }\left( \cos C_{nl}\tau \right) \frac{\left[
A_{n+2k}^{2}\cos \left( f_{n+2k}\tau \right) +B_{n+2k}^{2}\right] }{%
A_{n+2k}^{2}+B_{n+2k}^{2}}\left\vert 2,2\right\rangle \left\vert
1,-1\right\rangle \left\vert n-k,n-l\right\rangle  \notag \\
&&-i\sqrt{\left( 1-\alpha \right) }\left( \sin C_{nl}\tau \right)
A_{n+2k}B_{n+2k}\frac{\left[ \cos \left( f_{n+2k}\tau \right) -1\right] }{%
A_{n+2k}^{2}+B_{n+2k}^{2}}\left\vert 2,-2\right\rangle \left\vert
1,1\right\rangle \left\vert n-k+2,n-l-1\right\rangle  \notag \\
&&-\sqrt{\left( 1-\alpha \right) }\left( \sin C_{nl}\tau \right) A_{n+2k}%
\frac{\sin \left( f_{n+2k}\tau \right) }{\sqrt{\left(
A_{n+2k}^{2}+B_{n+2k}^{2}\right) }}\left\vert 2,0\right\rangle \left\vert
1,1\right\rangle \left\vert n-k+1,n-l-1\right\rangle  \notag \\
&&-i\sqrt{\left( 1-\alpha \right) }\left( \sin C_{nl}\tau \right) \frac{%
\left[ A_{n+2k}^{2}\cos \left( f_{n+2k}\tau \right) +B_{n+2k}^{2}\right] }{%
A_{n+2k}^{2}+B_{n+2k}^{2}}\left\vert 2,2\right\rangle \left\vert
1,1\right\rangle \left\vert n-k,n-l-1\right\rangle ,
\end{eqnarray}%
where

\begin{eqnarray}
A_{nk} &=&\sqrt{\left( n-k-1\right) },\;\;B_{nk}=\sqrt{\left( n-k\right) }%
,\;\;C_{nl}=\sqrt{\left( n-l\right) },  \notag \\
A_{n+2k} &=&\sqrt{\left( n-k+1\right) },\;\;B_{n+2k}=\sqrt{\left(
n-k+2\right) },  \notag \\
f_{nk} &=&\sqrt{2\left( A_{nk}^{2}+B_{nk}^{2}\right) },\ \ \ \ f_{n+2k}=%
\sqrt{2\left( A_{n+2k}^{2}+B_{n+2k}^{2}\right) }.
\end{eqnarray}

\label{A2}

The analytic form of $\left\vert \Phi _{A_{1}A_{2}B}^{n-k+1,n-l}(\tau
)\right\rangle $ in Eq. (\ref{ro2tau}) is given by 
\begin{eqnarray}
&&\left\vert \Phi _{A_{1}A_{2}B}^{n-k+1,n-l}(\tau )\right\rangle  \notag \\
&=&-i\cos \left( C_{nl}\tau \right) B_{n+1k}\frac{\sin \left( f_{n+1k}\tau
\right) }{\sqrt{\left( A_{n+1k}^{2}+B_{n+1k}^{2}\right) }}\left\vert
2,-2\right\rangle \left\vert 1,-1\right\rangle \left\vert
n-k+1,n-l\right\rangle  \notag \\
&&+\cos \left( C_{nl}\tau \right) \cos \left( f_{n+1k}\tau \right)
\left\vert 2,0\right\rangle \left\vert 1,-1\right\rangle \left\vert
n-k,n-l\right\rangle  \notag \\
&&-i\cos \left( C_{nl}\tau \right) A_{n+1k}\frac{\sin \left( f_{n+1k}\tau
\right) }{\sqrt{\left( A_{n+1k}^{2}+B_{n+1k}^{2}\right) }}\left\vert
2,2\right\rangle \left\vert 1,-1\right\rangle \left\vert
,n-k-1,n-l\right\rangle  \notag \\
&&-\sin \left( C_{nl}\tau \right) B_{n+1k}\frac{\sin \left( f_{n+1k}\tau
\right) }{\sqrt{\left( A_{n+1k}^{2}+B_{n+1k}^{2}\right) }}\left\vert
2,-2\right\rangle \left\vert 1,1\right\rangle \left\vert
n-k+1,n-l-1\right\rangle  \notag \\
&&-i\sin \left( C_{nl}\tau \right) \cos \left( f_{n+1k}\tau \right)
\left\vert 2,0\right\rangle \left\vert 1,1\right\rangle \left\vert
n-k,n-l-1\right\rangle  \notag \\
&&-\sin \left( C_{nl}\tau \right) A_{n+1k}\frac{\sin \left( f_{n+1k}\tau
\right) }{\sqrt{\left( A_{n+1k}^{2}+B_{n+1k}^{2}\right) }}\left\vert
2,2\right\rangle \left\vert 1,1\right\rangle \left\vert
n-k-1,n-l-1\right\rangle .
\end{eqnarray}

\end{document}